\documentclass[superscriptaddress, reprint]{revtex4-1}
\usepackage{hyperref}
\usepackage{graphicx}
\usepackage{gensymb}
\usepackage{dcolumn}
\usepackage{xcolor}
\usepackage{amssymb}
\begin{document}

\title{Competing electronic instabilities in the quadruple perovskite manganite PbMn$_{7}$O$_{12}$}

\author{R. D. Johnson}
\email{roger.johnson@ucl.ac.uk}
\affiliation{Department of Physics and Astronomy, University College London, Gower Street, London, WC1E 6BT United Kingdom}
\author{D. D. Khalyavin}
\author{P. Manuel}
\affiliation{ISIS facility, Rutherford Appleton Laboratory-STFC, Chilton, Didcot, OX11 0QX, United Kingdom}
\author{A. A. Belik}
\affiliation{International Center for Materials Nanoarchitectonics (WPI-MANA), National Institute for Materials Science (NIMS), Namiki 1-1, Tsukuba, Ibaraki 305-0044, Japan}

\date{\today}

\begin{abstract}
Structural behaviour of PbMn$_{7}$O$_{12}$ has been studied by high resolution synchrotron X-ray powder diffraction. This material belongs to a family of quadruple perovskite manganites that exhibit an incommensurate structural modulation associated with an orbital density wave. It has been found that the structural modulation in PbMn$_{7}$O$_{12}$ onsets at 294 K with the incommensurate propagation vector $\mathbf{k}_s=(0,0,\sim2.08)$. At 110 K another structural transition takes place where the propagation vector suddenly drops down to a \emph{quasi}-commensurate value $\mathbf{k}_s=(0,0,2.0060(6))$. The \emph{quasi}-commensurate phase is stable in the temperature range of 40K - 110 K, and below 40 K the propagation vector jumps back to the incommensurate value $\mathbf{k}_s=(0,0,\sim2.06)$. Both low temperature structural transitions are strongly first order with large thermal hysteresis. The orbital density wave in the \emph{quasi}-commensurate phase has been found to be substantially suppressed in comparison with the incommensurate phases, which naturally explains unusual magnetic behaviour recently reported for this perovskite. Analysis of the refined structural parameters revealed that that the presence of the \emph{quasi}-commensurate phase is likely to be associated with a competition between the Pb$^{2+}$ lone electron pair and Mn$^{3+}$ Jahn-Teller instabilities.
\end{abstract}

\maketitle

\section{Introduction}\label{SEC::intro}

The Jahn-Teller (JT) effect in transition metal compounds presents a direct interaction between the electronic configuration of the transition metal ions and the crystal structure. As such, it can mediate long-range ordering and cross-coupling of orbital, charge, and magnetic degrees of freedom \cite{Kugel82,Goodenough98} as exemplified by the simple perovskite manganites with generic chemical formula $A_{1-x}A'_{x}B$O$_{3}$ ($B$ = Mn) \cite{Goodenough55}. For example, in La$_{0.5}$Ca$_{0.5}$MnO$_3$ JT-active Mn$^{3+}$ and non-JT-active Mn$^{4+}$ ions adopt a 1:1 checkerboard type charge order. All Mn$^{3+}$ octahedral oxygen coordinations coherently elongate according to their JT instability, giving rise to long-range ordering of the Mn$^{3+}$ $d_{3z^2-r^2}$ orbitals. These orbitals in turn mediate ferromagnetic super-exchange interactions within zigzag chains, which then couple antiferromagnetically to form a so-called CE-type magnetic structure \cite{Koehler55,Radaelli97}.

A more complex pattern of magneto-orbital order preceded by charge order has been found in the quadruple perovskite manganites with generic chemical formula $AA'_3B_4$O$_{12}$ ($A' = B = \mathrm{Mn}$). For example, in CaMn$_7$O$_{12}$ 3:1 charge ordering of $B$ site Mn$^{3+}$ and Mn$^{4+}$ ions leads to a trigonal structural distortion that compresses the JT-active Mn$^{3+}$ octahedra, leading to an apparent ordering of $d_{x^2-y^2}$ orbitals \cite{bochu80}. However, anharmonic contributions to the elastic energy favour elongated rather than compressed Mn$^{3+}$ octahedra \cite{Khomskii00}, which results in an incommensurate orbital modulation (orbital density wave) that propagates along the trigonal axis with wavevector $\mathbf{k}_s=(0,0,k_z)$. The orbital density wave is characterised by an oscillation in the orbital mixing between $d_{3x^2-r^2}$ and $d_{3y^2-r^2}$, via $d_{x^2-y^2}$ \cite{Perks12}. Using the superspace formalism, this modulated state is described by the $R\bar{3}(00\gamma)0$ symmetry \cite{Slawinski09,Perks12}. At lower temperatures the incommensurate orbital modulation is accompanied by an incommensurate magnetic structure with a fundamental propagation vector, $\mathbf{k}_0$, that initially locks into the periodicity of the orbital modulation with $\mathbf{k}_0 = \mathbf{k}_s/2$, and then delocks at a second magnetic phase transition below which the system establishes its ground state \cite{Slawinski10,Johnson12,Slawinski12}. In the delocked ground state the magnetic structure is a constant moment helix with an incommensurately modulated spin chirality evidenced by higher order magnetic diffraction peaks with propagation vectors  $\mathbf{k}_{n+}=n\mathbf{k}_s + \mathbf{k}_0$ and $\mathbf{k}_{n-}=n\mathbf{k}_s - \mathbf{k}_0$, where $n=1,2,3...\infty$ \cite{Johnson16}.  This type of magnetic structure is stabilized by a set of competing exchange interactions and magnetic anisotropies modulated by the orbital density wave --- an effect referred to as magneto-orbital coupling.

The substitution of trivalent Bi for divalent Ca in BiMn$_7$O$_{12}$ makes all $B$ site Mn ions adopt a +3 oxidation state. Hence, the 3:1 charge ordering and the incommensurate orbital helix are eliminated and replaced by a commensurate JT-driven orbital order \cite{Mezzadri09,Belik17} analogous to that found in the simple perovskite LaMnO$_3$ \cite{RodriguezCarvajal98}. However, BiMn$_7$O$_{12}$ supports a second, polar structural instability associated with the Bi$^{3+}$ lone pair electrons, which gives rise to two, low symmetry ferroelectric phases not found in the divalent A site quadruple perovskites \cite{Mezzadri09,Slawinski17,Belik17}. Remarkably, an incommensurate orbital modulation can be established through light hole doping in BiCu$_{0.1}$Mn$_{6.9}$O$_{12}$ \cite{Khalyavin20}. In this doped material the incommensurate orbital order competes with the polar instabilities of the Bi$^{3+}$ lone pair electrons, which are forced to abandon their commensurate ferroelectric order and instead form the first example of a spontaneous incommensurate electric dipole helix \cite{Khalyavin20}.

In this paper we have brought together concepts established for CaMn$_7$O$_{12}$ and BiMn$_7$O$_{12}$, and applied them to PbMn$_7$O$_{12}$ --- a divalent $A$ site quadruple perovskite in which one might expect the competition between lone-pair and JT instabilities to play an important role in orbital and magnetic ordering. Indeed, the magnetic behaviour of this perovskite has been found to be substantially different from other members of the $A^{2+}$Mn$_7$O$_{12}$ family \cite{Johnson17}. Apart from the higher-temperature lock-in phase and the delocked ground state with modulated spin chirality, PbMn$_7$O$_{12}$ exhibits an additional intermediate magnetic phase in which the magnetic propagation vector is delocked from the structural modulation, but no modulation of the spin chirality was detected \cite{Johnson17}. As such, this intermediate phase appears to contradict the phenomenological model developed to explain magneto-orbital coupling in the $A^{2+}$Mn$_7$O$_{12}$ manganites \cite{Johnson16}.

We present a detailed crystallographic study of the low-temperature phases of PbMn$_7$O$_{12}$ in which we have discovered a new structural phase that spans both the lock-in magnetic phase and the intermediate magnetic phase. We demonstrate that this structural phase is characterised by an approximately commensurate structural modulation [$\mathbf{k}_s = (0,0,2.0060(6))$] that satisfies the lone pair instability of 2/3 Pb$^{3+}$ ions, but at considerable cost to the orbital order. These results reconcile the apparent discrepancy between the reported intermediate magnetic phase and the expected phenomenology. Furthermore, they present a scenario in which the competition between lone pair and JT instabilities is intimately linked to active charge degrees of freedom that may be tunable by appropriate doping schemes. Hence, PbMn$_7$O$_{12}$ brings fresh paradigms to the rich magneto-orbital physics of the divalent $A$ site quadruple perovskite manganites.

\section{Experiment}\label{SEC::exp}

A polycrystalline sample of PbMn$_7$O$_{12}$ was prepared from stoichiometric mixtures of Mn$_2$O$_3$ (99.997\%), MnO$_{1.839}$ (Alpha Aesar MnO$_2$ 99.997\% with the precise oxygen content determined by thermogravimetric analysis), and PbO (99.999\%). The mixtures were placed in Au capsules and treated at 6 GPa and 1373 K for 2 h (the duration of heating to the desired temperatures was 10 min) in a belt-type high-pressure apparatus. After the heat treatments, the samples were quenched to room temperature, and the pressure was slowly released. Synchrotron x-ray powder diffraction data were collected using the CRISTAL beamline at SOLEIL, the French national synchrotron facility. A fine polycrystalline sample of PbMn$_7$O$_{12}$ was loaded into a 0.3 mm diameter Lindemann capillary. The capillary was then mounted within a $^4$He flow cryostat installed on a 2-circle, high resolution powder diffractometer. Diffraction patterns were collected on cooling from 290 K down to a base temperature of 5 K, and on warming from 5 K up to 300 K. An x-ray wavelength of $0.58147 \mathrm{\AA}$ was selected for all measurements. Structural refinements were performed using \textsc{jana2006} \cite{Petricek14}.

\section{Results}\label{SEC::results}

At high temperature PbMn$_7$O$_{12}$ adopts a cubic crystal structure (space group $Im\bar{3}$) common to other divalent $A$ site quadruple perovskites \cite{Locherer12}. In this structure, Pb$^{2+}$ ions are 12-fold coordinated at the $A$ site, Mn$^{3+}$ ions are located in square-planar coordinations at the $A'$ site (labeled Mn1), and all $B$ sites are symmetry equivalent and occupied by Mn with an average oxidation state of $+3.25$. A structural phase transition associated with a 3:1 $B$ site charge ordering of Mn$^{3+}$ (labeled Mn2) and Mn$^{4+}$ (labeled Mn3), respectively, was reported to occur in the temperature range 380 - 397 K \cite{Locherer12,Belik16b}. The charge ordering is accompanied by a trigonal distortion of the crystal structure (space group $R\bar{3}$) and compression of the JT active B site Mn$^{3+}$ octahedra \cite{Locherer12,Belik16b}. This $R\bar{3}$ crystal structure was refined against our synchrotron powder diffraction data measured at 300 K and was found to be in excellent agreement. The refined structural parameters are given in Table \ref{TAB::crystalav}. Two trace impurity phases were identified as $\alpha$-Mn$_2$O$_3$ (2.0 wt\%) and Pb$_3$(CO$_3$)$_2$(OH)$_2$ (0.7 wt\%).

\begin{table}
\caption{\label{TAB::crystalav}$R\bar{3}$ crystal structure parameters of PbMn$_7$O$_{12}$ refined at 300 K, 160 K, 100 K, and 20 K. Below $T_\mathrm{OO1}$, these parameters correspond to the average, unmodulated structure. Atomic Wyckoff positions are as follows. Pb: $3a$ $[0,0,0]$, Mn1: $9e$ $[\frac{1}{2},0,0]$, Mn2: $9d$ $[\frac{1}{2},0,\frac{1}{2}]$, Mn3: $3b$ $[0,0,\frac{1}{2}]$, O1 \& O2: $18f$ $[x,y,z]$. $U_\mathrm{iso}$ is given in units of $\mathrm{\AA}^2$.}
\begin{ruledtabular}
\begin{tabular}{c | c c c c}
$T$ (K) & 300 & 160 & 100 & 20 \\
\hline
\multicolumn{5}{l}{\textbf{Lattice parameters}}\\
$a$ ($\mathrm{\AA}$)   & 10.5251(3) & 10.5138(2) & 10.5099(2) & 10.5087(2) \\
$c$ ($\mathrm{\AA}$)   & 6.4121(1) & 6.4109(1) & 6.4115(2) & 6.4094(1) \\
\hline
\multicolumn{5}{l}{\textbf{Fractional coordinates and a.d.p.s}}\\
Pb  \hspace*{\fill} $U_\mathrm{iso}$ & 0.00651(7) & 0.00651(6) & 0.00521(6) & 0.00405(5) \\
\\
Mn1 \hspace*{\fill} $U_\mathrm{iso}$ & 0.0065(1)  & 0.0056(2) & 0.0057(2) & 0.0042(2) \\
\\
Mn2 \hspace*{\fill} $U_\mathrm{iso}$ & 0.0027(1)  & 0.0041(1) & 0.0034(2) & 0.0033(1) \\
\\
Mn3 \hspace*{\fill} $U_\mathrm{iso}$ & 0.0030(3)  & 0.0048(3) & 0.0044(3) & 0.0042(3) \\
\\
O1  \hspace*{\fill} $x$ & 0.2269(4) & 0.2274(3) & 0.2271(3) & 0.2272(3)  \\
    \hspace*{\fill} $y$ & 0.2815(4) & 0.2792(3) & 0.2784(4) & 0.2786(3) \\
    \hspace*{\fill} $z$ & 0.0796(4) & 0.0819(4) & 0.0822(4) & 0.0817(4) \\
    \hspace*{\fill} $U_\mathrm{iso}$ & 0.0052(7) & 0.0041(7) & 0.0048(7) & 0.0037(6) \\
    \\
O2  \hspace*{\fill} $x$ & 0.3433(3) & 0.3417(3) & 0.3418(3) & 0.3419(3) \\
    \hspace*{\fill} $y$ & 0.5217(3) & 0.5208(3) & 0.5212(3) & 0.5207(3) \\
    \hspace*{\fill} $z$ & 0.3346(5) & 0.3339(4) & 0.3350(5) & 0.3347(4) \\
    \hspace*{\fill} $U_\mathrm{iso}$ & 0.0011(6) & 0.0037(8) & 0.0045(8) & 0.0035(7) \\
\hline
\multicolumn{5}{l}{\textbf{Reliability parameters (main reflections)}}\\
$R$ (\%)  & 2.5 & 1.1 & 1.1 & 0.9 \\
$wR$ (\%) & 3.0 & 1.4 & 1.4 & 1.2 \\
\end{tabular}
\end{ruledtabular}
\end{table}

\begin{figure}
\centering
\includegraphics[width=8.5cm]{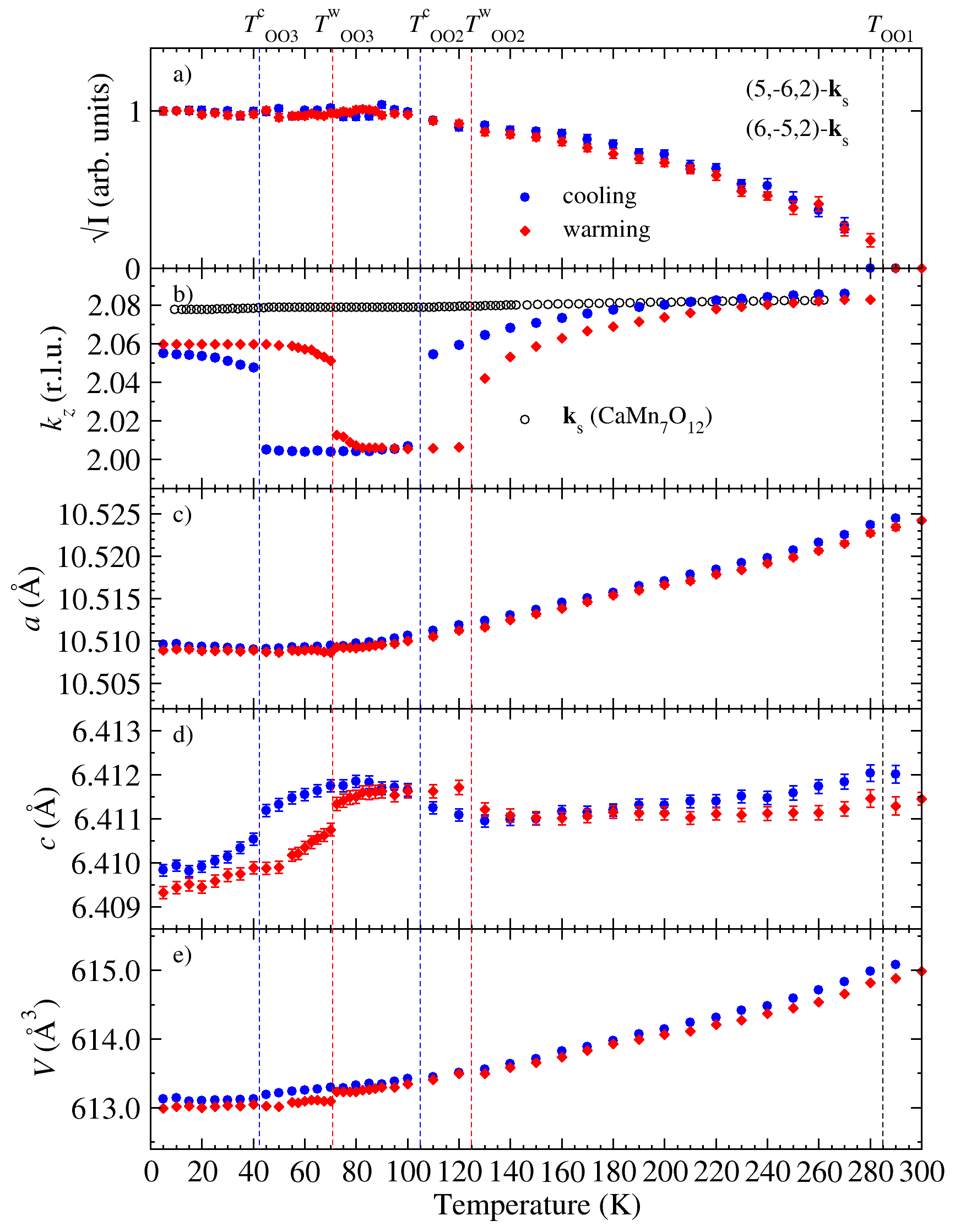}
\caption{\label{FIG::latt}Temperature dependence of a) the square root of the superimposed $(5,\bar{6},2)-\mathbf{k}_\mathrm{s}$ and $(6,\bar{5},2)-\mathbf{k}_\mathrm{s}$ integrated satellite intensities, b) the $z$ component of $\mathbf{k}_\mathrm{s}$, c) the $a$ lattice parameter, d) the $c$ lattice parameter, and e) the unti cell volume. Values refined on cooling and warming are shown as blue circles and red diamonds, respectively.}
\end{figure}

\begin{figure}
\centering
\includegraphics[width=8.5cm]{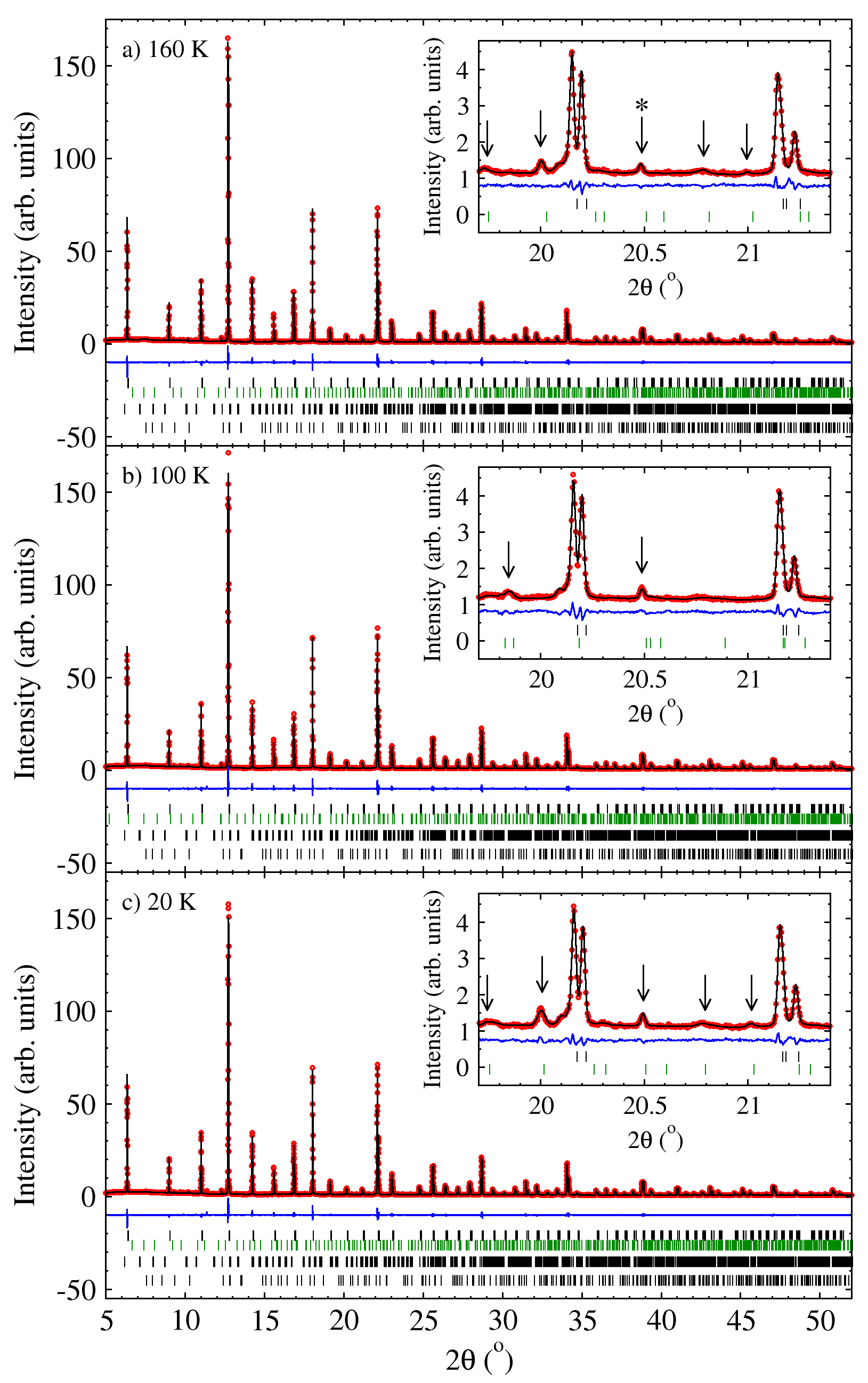}
\caption{\label{FIG::SXPDdata}Synchrotron x-ray powder diffraction data (red circles) measured at a) 160 K, b) 100 K, and c) 20 K. Fits of superspace structural models are shown as black lines, and the differences between $I_\mathrm{obs}$ and $I_\mathrm{calc}$ are given as blue lines. From top to bottom, the black tick marks show peak positions from the average PbMn$_7$O$_{12}$ structure, and the $\alpha$-Mn$_2$O$_3$ and Pb$_3$(CO$_3$)$_2$(OH)$_2$ impurities. The positions of incommensurate satellites originating in the structural modulation of PbMn$_7$O$_{12}$ are indicated by green tick marks, and highlighted in the insets by black arrows. The asterisk in the inset to (a) denotes the superimposed $(5,\bar{6},2)-\mathbf{k}_\mathrm{s}$ and $(6,\bar{5},2)-\mathbf{k}_\mathrm{s}$ peaks.}
\end{figure}

A second structural phase transiton at $T_\mathrm{OO1}=294~\mathrm{K}$ was also reported, and was assigned to the onset of orbital order \cite{Belik16b}. Below $T\sim T_\mathrm{OO1}$, a large number of weak peaks appeared in our diffraction data. These peaks could be indexed as satellite reflections of the main structural Bragg peaks, reached by the incommensurate propagation vector $\mathbf{k}_s = (0,0,k_z)$. The square root of the superimposed $(5,\bar{6},2)-\mathbf{k}_\mathrm{s}$ and $(6,\bar{5},2)-\mathbf{k}_\mathrm{s}$ integrated satellite intensities (marked by an asterisk in the inset to Figure \ref{FIG::SXPDdata}a) is shown in Figure \ref{FIG::latt}a, which demonstrates that an incommensurate structural modulation persisted from $T_\mathrm{OO1}$ down to 5 K.

The temperature dependence of $k_z$ is shown in Figure \ref{FIG::latt}b, alongside the temperature dependence of the same propagation vector component determined for CaMn$_7$O$_{12}$ (reproduced from reference \citenum{Johnson16}). Immediately below $T_\mathrm{OO1}$ both PbMn$_7$O$_{12}$ and CaMn$_7$O$_{12}$ supported structural modulations with $\mathbf{k}_s = (0,0,\sim2.08)$. However, on cooling PbMn$_7$O$_{12}$ below $T_\mathrm{OO2}^\mathrm{c} = 105~\mathrm{K}$ (superscript denotes cooling) there occurred a striking departure from the structural behaviour of CaMn$_7$O$_{12}$, marked by a downward step in $k_z$ to the \emph{quasi}-commensurate position $\mathbf{k}_s = (0,0,2.0060(6))$ (Figure \ref{FIG::latt}b). We note that the resolution of our diffraction experiment was sufficient to observe a significant difference from the commensurate vector $(0,0,2)$. The step in $k_z$ at $T_\mathrm{OO2}^\mathrm{c}$ was accompanied by an anomalous increase in the $c$ lattice parameter (Figure \ref{FIG::latt}d), indicating a significant strain-coupling between the structural modulation and the lattice. This \emph{quasi}-commensurate structural phase was found to persist down to $T_\mathrm{OO3}^\mathrm{c} = 43~\mathrm{K}$, at which point $k_z$ and $c$ stepped back towards values measured above $T_\mathrm{OO2}$ . On warming, a large thermal hysteresis of width 28 and 20 K was observed at $T_\mathrm{OO3}$ and $T_\mathrm{OO2}$, respectively (Figure \ref{FIG::latt}).

Microscopic details of the structural modulation in all three incommensurate phases ($T<T_\mathrm{OO3}$, $T_\mathrm{OO3}<T<T_\mathrm{OO2}$, and $T_\mathrm{OO2}<T<T_\mathrm{OO1}$) were established through refinement of a structural model defined within a 4D superspace formalism. The incommensurate atomic displacements of a given site were described by modulation amplitudes $\alpha_i$ and $\beta_i$, such that the atomic fractional coordinates in a unit cell reached by the lattice vector $[l_x,l_y,l_z]$ are
\begin{eqnarray}
x &=& x_0 + \alpha_x \sin(2\pi x_4) + \beta_x\cos(2\pi x_4) + l_x \nonumber\\
y &=& y_0 + \alpha_y \sin(2\pi x_4) + \beta_y\cos(2\pi x_4) + l_y \nonumber\\
z &=& z_0 + \alpha_z \sin(2\pi x_4) + \beta_z\cos(2\pi x_4) + l_z 
\end{eqnarray}
where $x_0$, $y_0$, and $z_0$ are the fractional coordinates in the average, unmodulated structure, and $x_4 = k_z\cdot(z_0+l_z)$. The modulation amplitudes and the average crystal structure parameters were constrained within the $R\bar{3}(00\gamma)0$ superspace group symmetry and refined against diffraction data measured at 160, 100, and 20 K, as shown in Figure \ref{FIG::SXPDdata}, which were chosen to represent the three incommensurate phases.
The superspace group was selected based on pyrocurrent \cite{Belik16a} and neutron powder diffraction measurements \cite{Johnson17}. The former revealed no change in electric polarization at the transition from paramagnetic to the magnetically ordered state, indicating that the paramagnetic space group was non-polar. The latter showed behaviour consistent with coupling between the magnetic subsystem and an orbital density wave associated with the $R\bar{3}(00\gamma)0$ symmetry \cite{Slawinski09,Slawinski10,Johnson16}. Excellent goodness of fit parameters were achieved, and the refined parameters and reliabilty factors are given in Tables \ref{TAB::crystalav} and \ref{TAB::crystalmod}.

\begin{table}
\caption{\label{TAB::crystalmod}Modulation amplitudes (in units $\mathrm{\AA}$) and refinement reliability parameters for the $R\bar{3}(00\gamma)0$ crystal structure of PbMn$_7$O$_{12}$ at 160 K, 100 K, and 20 K. Amplitudes that are zero by symmetry are not tabulated.}
\begin{ruledtabular}
\begin{tabular}{c | c c c}
$T$ (K) & 160 & 100 & 20 \\
\hline
\multicolumn{4}{l}{\textbf{Modulation amplitudes}}\\
Pb  \hspace*{\fill} $\alpha_z$ &-0.0100(2) &-0.0103(2) &-0.0144(1) \\
\\
Mn1 \hspace*{\fill} $\alpha_x$ & 0.0100(4) & 0.0119(4) & 0.0112(3) \\
    \hspace*{\fill} $\alpha_y$ &-0.0040(5) &-0.0004(7) &-0.0046(4) \\
    \hspace*{\fill} $\alpha_z$ & 0.0063(7) &-0.0021(6) & 0.0082(5) \\
\\
Mn2 \hspace*{\fill} $\alpha_x$ & 0.0029(4) & 0.0016(5) & 0.0034(3) \\
    \hspace*{\fill} $\alpha_y$ &-0.0015(5) &-0.0049(7) &-0.0034(4) \\
    \hspace*{\fill} $\alpha_z$ & 0.0030(8) &-0.0012(5) & 0.0029(6) \\
\\
Mn3 \hspace*{\fill} $\alpha_z$ &-0.001(1) &-0.002(1) &-0.0056(7) \\
\\
O1  \hspace*{\fill} $\alpha_x$ &-0.006(1) &-0.005(2) &-0.008(1) \\
    \hspace*{\fill} $\alpha_y$ & 0.004(1) & 0.001(1) & 0.0036(9) \\
    \hspace*{\fill} $\alpha_z$ & 0.004(2) & 0.001(2) &-0.001(2)\\
    \hspace*{\fill} $\beta_x$  &-0.004(1) &-0.002(2) &-0.003(1) \\
    \hspace*{\fill} $\beta_y$  & 0.000(1) &-0.003(2) & 0.000(1) \\
    \hspace*{\fill} $\beta_z$  & 0.007(2) &-0.003(2) & 0.007(2) \\
\\
O2  \hspace*{\fill} $\alpha_x$ & 0.004(1) & 0.003(2) & 0.009(1) \\
    \hspace*{\fill} $\alpha_y$ & 0.000(1) & 0.003(2) & 0.002(1) \\
    \hspace*{\fill} $\alpha_z$ & 0.014(2) &-0.010(2) & 0.009(2) \\
    \hspace*{\fill} $\beta_x$  &-0.006(1) & 0.001(2) &-0.004(1) \\
    \hspace*{\fill} $\beta_y$  & 0.000(1) &-0.001(2) &-0.001(1) \\
    \hspace*{\fill} $\beta_z$  & 0.008(1) & 0.011(2) & 0.012(1) \\
\hline
\multicolumn{4}{l}{\textbf{Reliability parameters (satellite reflections)}}\\
$R$ (\%)  & 6.8 & 5.9 & 4.8  \\
$wR$ (\%) & 4.9 & 4.3 & 4.2 \\
\end{tabular}
\end{ruledtabular}
\end{table}

\begin{figure}
\centering
\includegraphics[width=8.5cm]{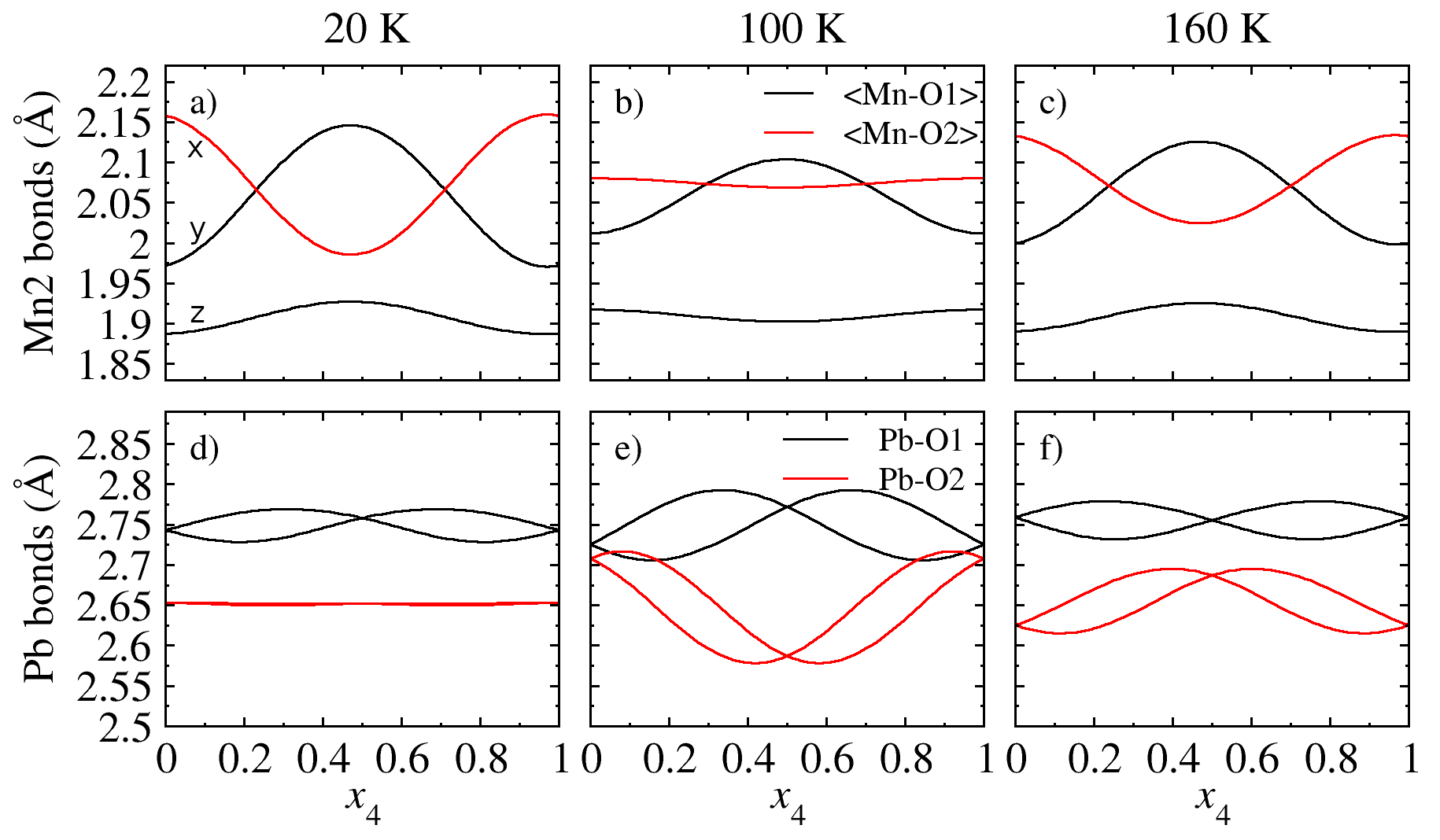}
\caption{\label{FIG::bonds}a)-c) Averaged Mn2-O bond length pairs and d)-f) Pb-O bond lengths, plotted as a function of $x_4$, calculated from the modulated structures refined at 20, 100, and 160 K, respectively. Six symmetry equivalent Pb-O1 (black) and Pb-O2 (red) bonds form the 12-fold oxygen coordination. The modulation origins are set to the Mn2 and Pb positions given in Table \ref{TAB::crystalav}. $\mathsf{x}$, $\mathsf{y}$, and $\mathsf{z}$ define a local coordinate system with $\mathsf{z}\sim || c$.}
\end{figure}

The relationship between the refined incommensurate structural modulation and the JT  structural instability can be understood in terms of the Mn2-O bond lengths, which are plotted against $x_4$ in Figures \ref{FIG::bonds}a-c. At 20 K (Figure \ref{FIG::bonds}a), the Mn2-O bonds alternate between 2 long bonds along $\mathsf{x}$ and 4 short bonds along $\mathsf{y}$ and $\mathsf{z}$, and 2 long bonds along $\mathsf{y}$ and 4 short bonds along $\mathsf{x}$ and $\mathsf{z}$ (here $\mathsf{x}$, $\mathsf{y}$, and $\mathsf{z}$ define a local coordinate system with $\mathsf{z}\sim || c$). This behaviour is consistent with an oscillation in the orbital occupation between $d_{3x^2-r^2}$ and $d_{3y^2-r^2}$, via the $d_{x^2-y^2}$ orbital, as found in CaMn$_7$O$_{12}$ \cite{Perks12}. The same qualitative behaviour was observed at 160 K (Figures \ref{FIG::bonds}c), albeit with a reduced amplitude. A significant change in this orbital modulation was observed at 100 K (Figures \ref{FIG::bonds}b), within the \emph{quasi}-commensurate phase. Here, the orbital modulation associated with the JT instability had been greatly suppressed, especially along the Mn2-O2 bonds.

A similar analysis of Pb-O bond lengths plotted against $x_4$ (Figure \ref{FIG::bonds}d-e) elucidates the relationship between the incommensurate structural modulation and the lone electron pair structural instability. The splitting of like Pb-O bonds, as plotted in Figures \ref{FIG::bonds}d-f, is consistent with polar distortions of the PbO$_{12}$ coordination. It is apparent that this splitting is largest in the \emph{quasi}-commensurate phase (Figure \ref{FIG::bonds}e), moderate at 160 K (Figure \ref{FIG::bonds}f), and largely suppressed at 20 K (Figure \ref{FIG::bonds}d).

The local structural distortions associated with the JT and lone electron pair instabilities were quantified as follows. Based on the orbital mixing angle formalism proposed by Goodenough \cite{Goodenough63}, a normalised orbital polarisation, $\eta(x_4)$, was calculated using the equation

\begin{equation}
\tan\bigg(\frac{\pi}{3} \eta(x_4)\bigg) = \sqrt{3}\frac{\bar{\mathsf{x}}(x_4) - \bar{\mathsf{y}}(x_4)}{\bar{\mathsf{x}}(x_4) + \bar{\mathsf{y}}(x_4) - 2\bar{\mathsf{z}}(x_4) }
\end{equation}
where $\bar{\mathsf{x}}(x_4)$, $\bar{\mathsf{y}}(x_4)$, and $\bar{\mathsf{z}}(x_4)$ are the average lengths of opposite Mn-O bond pairs.  By this definition $\eta=0$ corresponds to occupation of the $d_{x^2-y^2}$ orbital, and $\eta=\pm1$ correspond to maximal occupation of the $d_{3x^2-r^2}$ and $d_{3y^2-r^2}$ orbitals, respectively. The Pb polarisation, $\zeta(x_4)$ was taken to be the difference between the centre of mass of the 12 Pb oxygen ligands at positions $\mathbf{r}_{\mathrm{O}_i}(x_4)$ and the central Pb position, $\mathbf{r}_\mathrm{Pb}(x_4)$;
\begin{equation}
\zeta(x_4) = \bigg|\frac{1}{12}\sum_{i=1}^{12} \mathbf{r}_{\mathrm{O}_i}(x_4) - \mathbf{r}_\mathrm{Pb}(x_4)\bigg|
\end{equation}
where $\mathbf{r}_{\mathrm{O}_i}(x_4)$ and $\mathbf{r}_\mathrm{Pb}(x_4)$ were defined in a Cartesian basis in units $\mathrm{\AA}$.

\begin{figure}
\centering
\includegraphics[width=8.5cm]{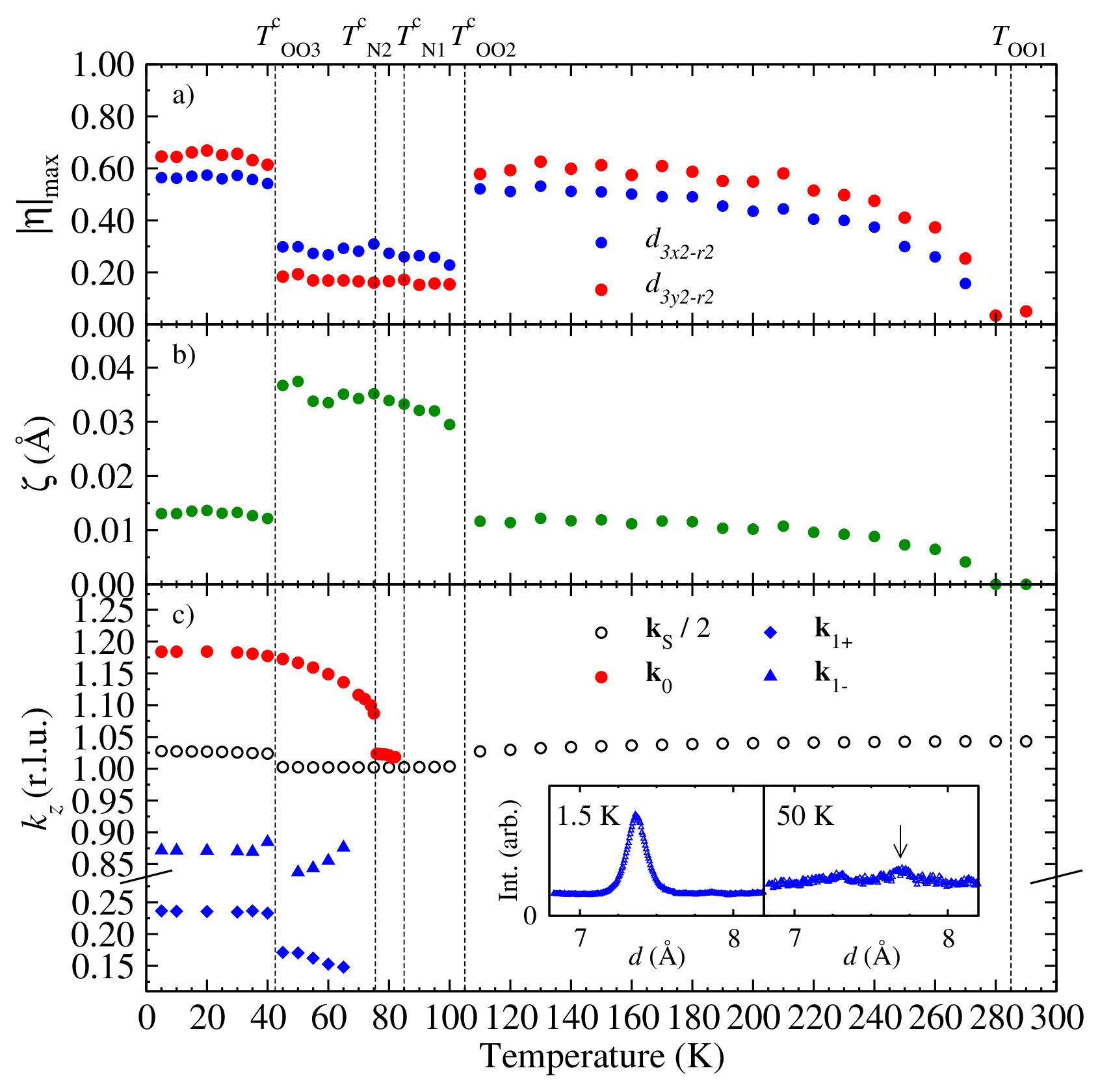}
\caption{\label{FIG::tempdep2}Temperature dependence of a) maximum values of the $d_{3x^2-r^2}$ and $d_{3y^2-r^2}$ normalised orbital polarisation, b) the Pb polarisation averaged over $x_4$, and c) the magnetic propagation vectors reproduced from reference \citenum{Johnson17} plotted alongside $\mathbf{k}_\mathrm{s}/2$. All values were determined from data measured on cooling. The inset to (c) highlights $\mathbf{k}_{1-}$ magnetic diffraction peaks measured at 1.5 K (left) and 50 K (right, marked by a black arrow), also reproduced from reference \citenum{Johnson17}.}
\end{figure}

Figure \ref{FIG::tempdep2}a shows the maximum absolute value of the $d_{3x^2-r^2}$ and $d_{3y^2-r^2}$ normalised orbital polarisations evaluated on cooling from 300 down to 5 K. The orbital polarisation was found to grow below $T_\mathrm{OO1}$, and reached a maximum value of 0.6 (60\% polarised) at the lowest measured temperature. This value is considerably reduced compared to the 95\% orbital polarisation observed in CaMn$_7$O$_{12}$ \cite{Perks12}, indicating partial suppression of the orbital order at all temperatures. The \emph{quasi}-commensurate phase between $T_\mathrm{OO3}$ and $T_\mathrm{OO2}$ is characterised by a large drop in both orbital polarisations, consistent with the change in bond lengths plotted in Figure \ref{FIG::bonds}. The opposite behaviour was observed in the temperature dependence of the Pb polarisation, which is shown averaged over $x_4$ in Figure \ref{FIG::tempdep2}b. A small Pb polarisation evolved below $T_\mathrm{OO1}$, which became greatly enhanced in the \emph{quasi}-commensurate phase.

These results demonstrate that the new \emph{quasi}-commensurate structural phase of PbMn$_7$O$_{12}$ arises due to competition between Mn2 JT and Pb lone-pair instabilities. The nature of this competition and the implications for long-range magnetic order are discussed in the following section.

\section{Discussion}\label{SEC::discussion}

\begin{figure}
\centering
\includegraphics[width=8.5cm]{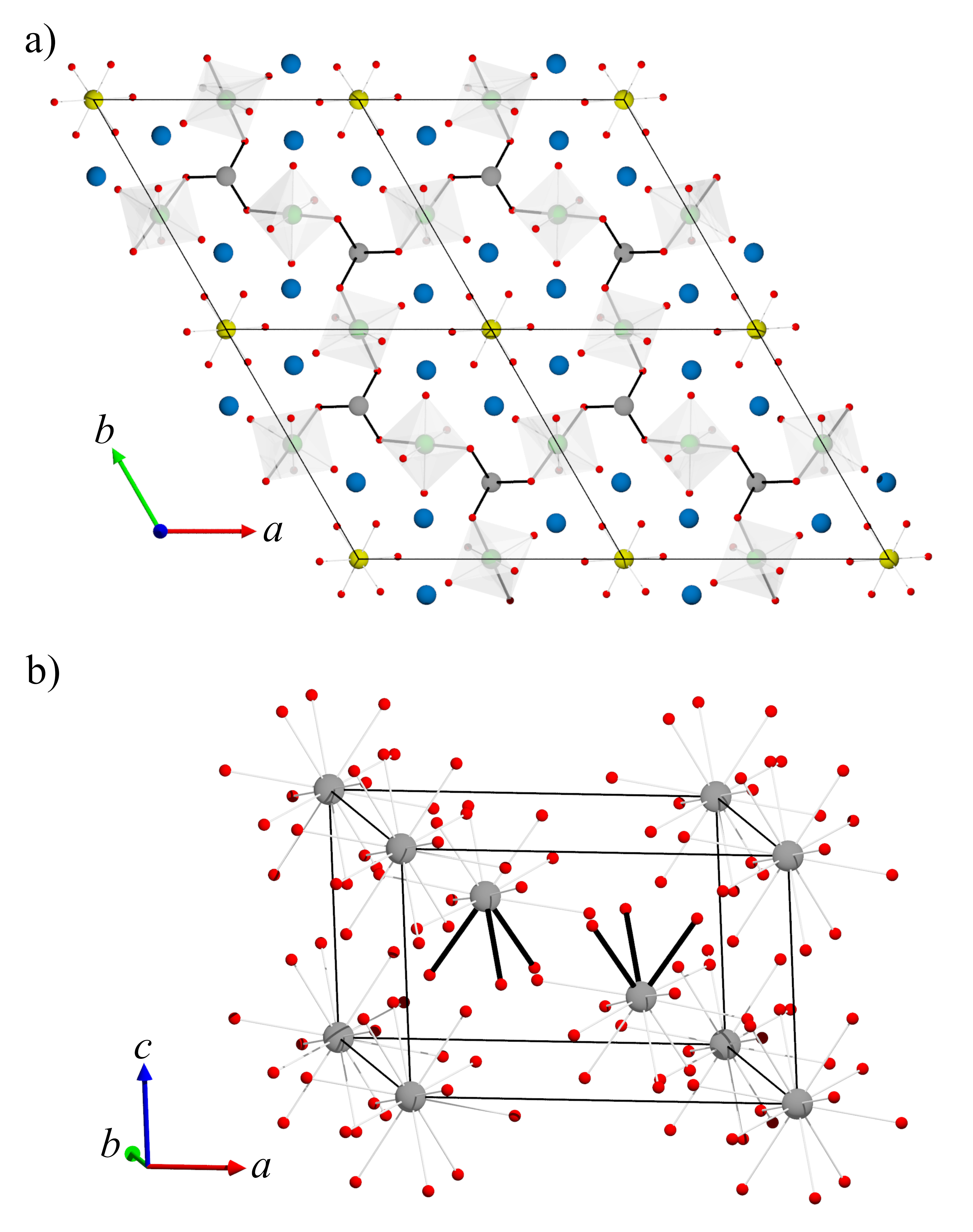}
\caption{\label{FIG::structure} The \emph{quasi}-commensurate crystal structure of PbMn$_7$O$_{12}$ drawn in the commensurate limit, in which the Pb lone-pair instability is maximally stabilised (see text). a) the layer of atoms centred around $z=1/2$, b) 3D view of PbO$_{12}$ coordinations. Pb, Mn1, Mn2, Mn3, and O atoms are drawn as grey, blue, green, yellow, and red spheres, respectively. JT active Mn2 octahedra are shaded grey. Short Pb-O bonds are highlighted in bold, and the commensureate unit cell is drawn as fine black lines.}
\end{figure}

The role of Pb lone-pair instabilities in establishing the \emph{quasi}-commensurate phase can be understood by considering the crystal structure in the commensurate limit; $\mathbf{k}_s = (0,0,2)$. The symmetry of this structure depends on the global phase of the modulation, and as such can be either centrosymmetric $P\bar{3}$ or non-centrosymmetric $P3$. In the former case, only sine terms are present ($\beta_i = 0$) while in the latter, an admixture of sine and cosine terms is allowed. Let us consider the more symmetric structure $P\bar{3}$. In this structure, two out of the three Pb sites related by $R$-centring translation of the parent structure obtain a large polarisation, while one site has zero polarisation. The Pb polarization is stabilized by the formation of asymmetric short Pb-O bonds. In the layer of atoms centred around $z=1/2$, short Pb-O bonds and Mn2 JT elongation axes form a coherent honeycomb motif (Figure \ref{FIG::structure}a) in which both structural instabilities are satisfied. However, in the layer of atoms centred around $z=0$ (not shown), both Pb lone-pair and Mn2 orbital polarisations are zero. Figure \ref{FIG::structure}b gives a 3D view of the commensurate unit cell that highlights the antiferroelectric arrangement of polar PbO$_{12}$ distortions at the two Pb sites in the centre of the commensurate unit cell. Assuming that in the layers with the antiferroelectric distortions the lone pair instability is satisfied, the absolute value of the local polarization, averaged over the period of the commensurate modulation, is $2/3 \approx 0.667$ of the hypothetical "fully polarised state". In the case of an incommensurate modulation, the averaged polarization is $2/\pi \approx 0.637$. Thus, the commensurate antiferrolectric structure better optimizes the lone pair electronic instability. 

Despite the favourable relationship between the lone pair instability and the commensurate structure, we find that the refined propagation vector is slightly off the commensurate value in the \emph{quasi}-commensurate phase. This experimental finding might be associated with the formation of an ordered pattern of commensurate translational domains with modulation (phase) slips at the domain boundaries. By this mechanism the system can systematically omit structural layers with zero polarisation and thereby increase the density of the layers shown in Figure \ref{FIG::structure}a. In this case the frequent repetition of the translational domains result in a discrete (step-like) modulated state that gives incommensurate satellite reflections in diffraction experiments. This is a well-known phenomenon for some alloys with so-called long period antiphase boundary modulated structures \cite{VanLanduyt85}. Furthermore, an anharmonic step-like incommensurate structural modulation was proposed for another perovskite system with a lone pair electronic instability; Bi$_{0.75}$La$_{0.25}$FeO$_{3}$ \cite{Rusakov11}.  To some extent, this model can also be considered as an ordered pattern of antiphase domains of a commensurate antipolar structure.

PbMn$_7$O$_{12}$ was reported to undergo three, low temperature magnetic phase transitions at $T_\mathrm{N1}=83~\mathrm{K}$, $T_\mathrm{N2}=77~\mathrm{K}$, and $T_\mathrm{N3}=43~\mathrm{K}$, with a large thermal hysteresis observed at $T_\mathrm{N3}$ between approx. 37 and 65 K \cite{Belik16a}. In the first magnetically ordered phase ($T_\mathrm{N2} < T < T_\mathrm{N1}$), PbMn$_7$O$_{12}$ was found to adopt the same lock-in magnetic structure as CaMn$_7$O$_{12}$ \cite{Johnson17}, albeit over a much narrower temperature range. The ground state magnetic structure of PbMn$_7$O$_{12}$ ($T < T_\mathrm{N3}$) was found to be a de-locked helix with modulated spin chirality (also observed in the ground state of CaMn$_7$O$_{12}$ \cite{Johnson16,Johnson17}); evidenced by $\mathbf{k}_{1+}$ and $\mathbf{k}_{1-}$ magnetic diffraction peaks. According to the phenomenological model of magneto-orbital coupling \cite{Johnson16}, the observation of a modulated spin chirality below $T_\mathrm{N3}$ implied the presence of an incommensurate orbital density wave with $R\bar{3}(00\gamma)0$ symmetry. This inference is now confirmed by the results discussed in Section \ref{SEC::results}.

Remarkably, in the intermediate magnetic phase of PbMn$_7$O$_{12}$ ($T_{N3}<T<T_{N2}$) the fundamental magnetic order was found to be delocked as in the ground state, but the $\mathbf{k}_{1+}$ and $\mathbf{k}_{1-}$ magnetic diffraction peaks associated with the magneto-orbital coupling were apparently missing. This result indicated that either the magnetic and orbital orders unexpectantly decoupled, or that the orbital modulation was temporarily destroyed \cite{Johnson17}. The above results demonstrate that the magnetic phase transition at $T_{N3}$ is in fact a magnetostructural phase transition with $T_{N3}=T_{OO3}$. This relationship is exemplified by the large thermal hysteresis observed at both $T_{N3}$ and $T_{OO3}$. Hence, the intermediate magnetic phase occurs due to the stabilisation of the concomitant \emph{quasi}-commensurate structural phase.

The phenomenological relationship between magnetic and structural propagation vectors that holds in the ground state, implies that the $\mathbf{k}_{1+}$ and $\mathbf{k}_{1-}$ magnetic peaks in the intermediate phase will have shifted position in accordance with the newly established \emph{quasi}-commensurate propagation vector. Furthermore, the large suppression of the orbital order in the \emph{quasi}-commensurate phase would lead to a suppression in the spin-chirality modulation and hence a reduction in the $\mathbf{k}_{1+}$ and $\mathbf{k}_{1-}$ diffraction intensities. Careful inspection of the neutron powder diffraction data reported in reference \citenum{Johnson17} (reproduced in the Figure \ref{FIG::tempdep2}c inset) showed the presence of weak diffraction peaks at the shifted $\mathbf{k}_{1-}$ position, close to the detection limit of the experiment. The temperature dependence of the measured $\mathbf{k}_{1-}$ and calculated $\mathbf{k}_{1+}$ (based upon the former) propagation vectors within the intermediate phase is shown in Figure \ref{FIG::tempdep2}c, alongside the other magnetic propagation vectors also reproduced from reference \citenum{Johnson17}. Thus, the unusual magnetic behaviour of PbMn$_{7}$O$_{12}$ can be naturally understood based on the above structural characterisation and the phenomenological model of the magneto-orbital coupling, previously developed to explain the lock-in phase and multi-\textbf{k} magnetic ground state in CaMn$_{7}$O$_{12}$ \cite{Johnson16}.

\section{Conclusions}\label{SEC::conc}

The quadruple perovskite manganite PbMn$_{7}$O$_{12}$ exhibits a set of structural phase transitions upon cooling below room temperature. The higher temperature transition at $T_\mathrm{OO1}=294 ~\mathrm{K}$ is associated with development of an orbital density wave with incommensurate propagation vector $\mathbf{k}_{s}=(0,0,\sim2.08)$. At $T_\mathrm{OO2}=110~\mathrm{K}$, the propagation vector suddenly changes taking the nearly commensurate value $\mathbf{k}_{s}=(0,0,2.0060(6))$. The \emph{quasi}-commensurate modulation holds down to $T_\mathrm{OO3}=40~\mathrm{K}$, when a re-entrant transition back to the incommensurate phase takes place. The orbital density wave is significantly suppressed in the \emph{quasi}-commensurate phase while the structural distortions optimizing the lone-pair electronic instability of Pb$^{2+}$ are enhanced. This behaviour is interpreted as a competition between the lone pair and JT instabilities. The magnetic order that emerges in the \emph{quasi}-commensurate phase reflects these structural changes specific to this perovskite. In particular, the puzzling absence of the magneto-orbital coupling and associated modulation of the spin-chirality reported in the intermediate magnetic phase ($40 < T < 65 ~\mathrm{K}$), is due to the severe reduction of the orbital density wave in the \emph{quasi}-commensurate regime. The re-entrant transition to the incommensurate structural phase re-establishes the orbital polarization which in turn triggers a strong magneto-orbital coupling in the ground state.

\begin{acknowledgments}
R.D.J. acknowledges support from a Royal Society University Research Fellowship. A.A.B. acknowledges partial support from JSPS KAKENHI (Grant Number JP20H05276), a research grant (40-37) from Nippon Sheet Glass Foundation for Materials Science and Engineering, and Innovative Science and Technology Initiative for Security (Grant Number JPJ004596) from Acquisition, Technology, and Logistics Agency (ATLA), Japan. We acknowledge SOLEIL for provision of synchrotron radiation facilities (Proposal No. 20171110) and we would like to thank E. Elkaim for assistance in using beamline Cristal.
\end{acknowledgments}

\bibliography{pmo}

\end{document}